\documentstyle[aps, amsmath, multicol, graphicx]{revtex}
\begin{document}

\draft

\title{Exciting, Cooling And Vortex Trapping In A Bose-Condensed Gas}
\author{R.J. Marshall, G.H.C. New}
\address{Laser Optics and Spectroscopy Group, Department of Physics, Imperial College, London \mbox{SW7 2BZ} United Kingdom.\\ http://www.lsr.ph.ic.ac.uk/LTG/index.html}
\author{K. Burnett, S. Choi}
\address{Clarendon Laboratory, Physics Department, University of Oxford, Oxford OX1 3PU England United Kingdom}
\date{\today}
\maketitle

\begin{abstract}
A straight forward numerical technique, based on the Gross-Pitaevskii equation, is used to generate a self-consistent description of thermally-excited states of a dilute boson gas.
The process of evaporative cooling is then modelled by following the time evolution of the system using the same equation.
It is shown that the subsequent rethermalisation of the thermally-excited state produces a cooler coherent condensate.
Other results presented show that trapping vortex states with the ground state may be possible in a two-dimensional experimental environment.
\end{abstract}

\begin{multicols}{2}
%%
%%
%% SECTION 1
\section{Introduction}
\label{sec:intro}

Demonstrations of Bose-Einstein condensation (BEC) in trapped, dilute alkali gases\cite{bec1,bec2,bec3} provide a strong incentive to develop accurate theoretical models of the phenomenon.
In particular, theories that go beyond the simplest approach based on the Gross-Pitaevskii (GP) equation \cite{gp} are under intense study \cite{t1,t2,ev1,ev2}.
In this paper, we show how the GP equation can handle the time evolution of Bose condensed gases at finite temperature or in non-equilibrium states.

Since current experiments produce BECs at a finite temperature, the resulting Bose-condensed vapour is, therefore, a mixture of condensate and excitations.
Understanding the excited states through the Gross-Pitaevskii equation however, has generally been limited to studying the linearised excitations of the system \cite{ex1,ex2,ex3,sam}.
It is possible, using finite temperature field theory, to calculate the energy and mean field due to the excited quasiparticles in a self consistent manner for static configurations \cite{hardcore}.
To study the time evolution of thermally-excited condensates using such techniques does however, require a prohibitive amount of computational time even for a one-dimensional system.
The result is that there is no computationally simple way to study the evolution of thermally-excited states or self-consistent combinations of them that employs the fuller quantum field theory approach.
This is, of course, a very great limitation in modelling experiments.

The evaporative cooling process \cite{ev1,ev2}, where hotter atoms are removed and the resulting rethermalisation of the atomic gas produces an overall cooling is, of course, fundamental to producing a BEC.
Current theoretical descriptions \cite{evt1,evt2,evt3,evt4,kin1,kin2} generally rely on the use of kinetic equations or approximations to them.
To date, evaporative cooling has not been studied using the GP equation, although it is implicit in the work of Kagan and Svistunov \cite{kegan,helpme,helpme2}.
We shall not be studying the problems associated with the critical point and we refer the reader to the recent work of Stoof, Ref. \cite{stoof}, for a discussion of related issues.

This paper addresses the evolution of a thermally-excited boson gas, including evaporative cooling, through a Gross-Pitaevskii formalism.
Section \ref{sec:back} contains an outline of the theoretical and numerical description of a Bose-condensed  atomic vapour based on the GP equation which is used throughout the following sections.

In section \ref{sec:exs} a numerically-efficient method for generating self-consistent thermally-excited states of a boson gas in three spatial dimensions is presented.
It is based on the GP equation which describes the thermalisation process or self-interaction of the mean field through the nonlinearity.
Results of a two-dimensional numerical simulation using parameters corresponding to a sample of $^{87}\text{Rb}$ atoms indicate that the method is robust and largely independent of the initial distribution

Finally, in section \ref{sec:evap}, a model of the evaporative cooling process based on the mean field description is presented.
In the model the most energetic atoms are removed from the atomic vapour, where neither the assumption of sufficient ergodicity nor approximations relating to the strength of the interactions between the thermal and condensate parts of the thermally-excited state are necessary.
Results from a numerical simulation of this model using a similar sample of atoms are also included.

%%
%%
%% SECTION 2
\section{Theoretical and Numerical Background}
\label{sec:back}

The evolution of the mean field, $\Psi({\boldsymbol{r}},t)$, is described by the Gross-Pitaevskii equation \cite{gp}
\begin{eqnarray}
\label{equ:gp}
i\hbar\partial_t \Psi({\boldsymbol{r}},t) & = & \frac{-\hbar^2\nabla^2}{2m}\Psi({\boldsymbol{r}},t) + V({\boldsymbol{r}})\Psi({\boldsymbol{r}},t) \\\nonumber 
& & + NU_0|\Psi({\boldsymbol{r}},t)|^2\Psi({\boldsymbol{r}},t)
\end{eqnarray}
where $m$ is the mass of the atoms and $N$ is the number of atoms in the atomic vapour.
The mean field is the ensemble average of the boson field operator, $\hat{\Psi}({\boldsymbol{r}},t)$, where the nature of the ensemble will be addressed below.
The harmonic trapping potential and nonlinear constant are defined by
\begin{mathletters}
\begin{eqnarray}
V({\boldsymbol{r}}) = m\omega_t{\boldsymbol{r}}\cdot{\boldsymbol{r}}/2\\
U_0 = 4\pi\hbar^2 a/m
\end{eqnarray}
\end{mathletters}
where $\omega_t$ is the angular frequency of the trapping potential and $a$ is the inter-atomic s-wave scattering length.
The self-interaction of the mean field, describing the two particle scattering of the atoms, is determined by the nonlinearity in Eq.\ (\ref{equ:gp}) and is characterised by the number of atoms and their relative scattering strengths.
The explicit appearance of the number of atoms in Eq.\ (\ref{equ:gp}) is associated with the normalisation condition
\begin{equation}
\int |\Psi({\boldsymbol{r}},t)|^2 {\mathbf{dr}} = 1
\end{equation}

To facilitate a computational solution, Eq.\ (\ref{equ:gp}) is scaled in terms of harmonic oscillator units
\begin{equation}
{\boldsymbol{r}} = {\boldsymbol{\zeta}}\left(\frac{\hbar}{2m\omega_t}\right)^{\frac{1}{2}},\quad t = \tau/\omega_t
\end{equation}
resulting in
\begin{eqnarray}
i\partial_\tau \Psi({\boldsymbol{\zeta}},\tau) & = & \left(-\nabla^2 + \frac{x^2+y^2+\epsilon z^2}{4} \right)\Psi({\boldsymbol{\zeta}},\tau) \\\nonumber
& & + \frac{NU_0}{\hbar\omega_t}|\Psi({\boldsymbol{\zeta}},\tau)|^2 \Psi({\boldsymbol{\zeta}},\tau)
\end{eqnarray}
where ${\boldsymbol{\zeta}} = (x,y,z)$ and $\epsilon$ is ratio of trapping potentials.

In a `Bose-Einstein membrane' the atomic de Broglie wave is confined strongly in one-dimension, so the ratio of the trapping potential, $\epsilon$, is large.
For this two dimensional, large $\epsilon$, scenario there will be no excitations in the third, confined, dimension due to the much larger energy that they would require.
Furthermore, in this confined dimension the number of atoms per unit length will be small allowing the mean field to be represented by a ground simple harmonic oscillator (SHO) state, namely
\begin{equation}
\label{equ:ap}
\Psi({\boldsymbol{\zeta}},\tau) = \Psi(x,y,\tau)\left(\frac{\epsilon}{4\pi^2}\right)^{1/4}\exp\left(\frac{-\epsilon^{1/2}z^2}{4}\right)
\end{equation}
Substituting Eq.\ (\ref{equ:ap}) into the GP equation results in
\begin{eqnarray}
i\partial_\tau \Psi({\boldsymbol{\xi}},\tau) & = & \left(-\nabla^2 + \frac{{\boldsymbol{\xi}}^2}{4}  \right)\Psi({\boldsymbol{\xi}},\tau) \\\nonumber
& & + \eta(z)|\Psi({\boldsymbol{\xi}},\tau)|^2\exp\left(\frac{-\epsilon^{1/2}z^2}{4}\right) \Psi({\boldsymbol{\xi}},\tau)
\end{eqnarray}
where ${\boldsymbol{\xi}} = (x,y)$. Here the nonlinear term, $\eta(z) = \frac{N U_0 \epsilon^{1/4}}{\hbar\omega_t(2\pi)^{1/2}}$, is parameterised by the confined dimension.
Choosing $z=0$ results in the following two dimensional GP equation,
\begin{equation}
\label{equ:gps}
i\partial_\tau \Psi({\boldsymbol{\xi}},\tau) = \left(-\nabla^2 + \frac{\xi^2}{4} + \eta |\Psi({\boldsymbol{\xi}},\tau)|^2\right)\Psi({\boldsymbol{\xi}},\tau),
\end{equation}
where $\eta = \eta(0) = \frac{N U_0 \epsilon^{1/4}}{\hbar\omega_t(2\pi)^{1/2}}$.
Here $\Psi({\boldsymbol{\xi}},\tau)$ is the mean field surface density, to reconstruct the full volume density one simply multiplies the surface density by the ground SHO state.

The formal solution of Eq.\ (\ref{equ:gps}) is given by
\begin{equation}
\label{equ:fs}
\Psi({\boldsymbol{\xi}},\tau+\delta \tau) = e^{-i \delta t\hat{H}}\Psi({\boldsymbol{\xi}},\tau)
\end{equation}
where $\delta t$ is the time step and $\hat{H} = (-\nabla^2 + \xi^2/4 + \eta|\Psi({\boldsymbol{\xi}},\tau)|^2)$.

Eq.\ (\ref{equ:fs}) allows a numerical solution using the split-step operator method.
The solution, performed in two dimensions, exhibits features that are extremely difficult to see in the one-dimensional models; it also enables the study of the crucial process of vortex formation.
Furthermore, this solution is a direct simulation of a two-dimensional `Bose-Einstein membrane', realisation of which in two-dimensional atomic waveguides is a subject of increasing interest, with recent experimental and theoretical advances \cite{2dbec1,2dbec2}.

It is pertinent to note that the analysis leading to Eq.\ (\ref{equ:gps}) represents an approximation of the available quasiparticle modes.
This approximation results in the phenomena presented occurring over longer time scales than would be expected in an experimental configuration.
Furthermore, the quantitative time scales for the two-dimensional and three-dimensional cases are, of course, quite distinct as the corresponding increase in mode density results in shorter time scales over which the phenomena presented occur.
Nevertheless the topics presented in this paper (self-consistent thermally-excited state generation, evaporative cooling and the mechanisms relating to them) are not specifically related to the number of dimensions considered.
Therefore, the arguments and discussion presented in this paper apply to the phenomenon of Bose-Condensation in general.

Results from numerical simulations presented in this paper were carried out using typical experimental values for a trap containing $^{87}\text{Rb}$ atoms.
The values are given in table \ref{tab:expt}.
\narrowtext
\begin{table}
\caption{Values of constants used in the numerical simulation.}
\label{tab:expt}
\begin{tabular}{dd}
Constant & Value\\
\tableline
$m$ & 1.4$\times$10$^{-25}$ kg\\
$a$ & 5.2$\times$10$^{-9}$ m\\
$\omega_t$ & 20.0$\times\pi$ rad s$^{-1}$
\end{tabular}
\end{table}

%%
%%
%% SECTION 3
\section{Self-Consistent Thermally-Excited State Generation}
\label{sec:exs}

To overcome the prohibitive amount of computational time required to time evolve a thermally-excited condensate using finite temperature field theory, an alternative technique has been devised.
This technique for generating a self-consistent thermally-excited state involves taking a noninteracting mean field and propagating it using the GP equation (Eq.\ (\ref{equ:gp})), whilst the interactions are slowly switched on.

One should expect the GP equation to describe the kinetics and dynamics of a finite temperature boson gas as long as the relevant modes of excitation are sufficiently well populated.
Indeed, Moore and Turok\cite{ewt} have shown that one can use the evolution of the classical field during the electroweak phase transition to determine the dynamics of the infrared modes of the quantum field.
The classical field description neglects quantum fluctuations (and other quantum manifestations, such as squeezed states or spontaneous emission).
The method works if, and only if, the population of the normal modes of the system are much greater than unity.
The normal modes of the system can then be treated as classical objects or quasiparticles that are statistically independent of the energy levels, allowing them to be represented by a classical field. Morgan et.\ al. \cite{sam} have also stressed that the use of a classical field to describe a finite temperature BEC is valid in the limit of large mode occupations.

The GP equation includes the dynamics and kinetics of a boson gas through the nonlinear term in the Hamiltonian, which in effect represents all the stimulated two-body scattering processes.
These processes ensure that an initial distribution will, given sufficient time, result in a thermal distribution of the quasiparticles, irrespective of the quantitative nature of the initial population distribution.
The resulting population distribution is close to the full Bose-Einstein distribution in the equipartition sense, i.e.
\begin{equation}
\frac{1}{e^{E_{nm}/k_BT}-1} \sim \frac{k_BT}{E_{nm}},
\end{equation}
where $k_B$ is the Boltzmann constant, $T$ is the temperature and $E_{nm}$ is the mode energy.
In this sense the quasiparticles act as classical objects, macroscopically populating the normal modes of the system.

If the GP equation provides an accurate description of the quasiparticle interactions, and therefore the time dependence of the boson gas, it must also describe its transition to equilibrium.
For example, a boson gas could be prepared with a non-thermal distribution of the energy level populations.
Given enough time however, the collisional interactions within the gas would rethermalise it, recreating a Bose-Einstein population distribution, in the equipartition sense.
Kagan et.\ al.\cite{kegan} argue that, for large quasiparticle populations, this is true even when the initial distribution is far from equilibrium or where the ground state is initially unpopulated.
This issue has however, been the subject of considerable debate \cite{stoof}.

To ensure that the resulting ensemble is fully representative and that the subsequent `nonlinear mixing' is unbiased, the initial distribution must include all symmetry combinations.
This requirement can be deduced from an explicit consideration of the parity invariance of the evolution of the SHO mode amplitudes.
The consequent symmetry dependent nonlinear mixing implies that if one symmetry combination is omitted the resulting nonlinear mixing will be unbalanced.
In particular, if the ensemble is comprised of entirely even or odd modes, resulting in a totally even nonlinear operator, the modes only couple between like symmetries producing a purely even or odd distribution.

The thermalisation or nonlinear mixing of the mode populations described by the nonlinear term in the Hamiltonian of th GP equation is the underlining process of this method of self-consistent thermally-excited state generation. It allows a direct solution of the finite temperature analysis to be circumvented when the quasiparticle populations are much greater than unity.
The self-consistent thermally-excited state generation is computationally achieved by taking a noninteracting, non-equilibrium, thermally-excited distribution and adiabatically increasing the nonlinear term.
Adiabatically switching on the interactions results in the relaxation of the mean field into an interacting, equilibrium distribution.

To this end the initial state is a thermally distributed superposition of SHO modes and therefore, a thermally-excited solution to the $N=0$ GP equation.
The initial state, $\Psi_{\mathrm{I}}({\boldsymbol{\xi}},\tau)$, is given by
\begin{equation}
\Psi_{\mathrm{I}}({\boldsymbol{\xi}},\tau) = \sum_{n,m}^{O,P}|a_{nm}|e^{i\alpha_{nm}}\psi_{nm}({\boldsymbol{\xi}},\tau)
\end{equation}
where $O,P$ are the mode-cut-off numbers, $|a_{nm}|$ is the modulus of the mode amplitude and $\alpha_{nm}$ is the mode phase. The SHO modes, $\psi_{nm}({\boldsymbol{\xi}},\tau)$, in Cartesian coordinates are given by
\begin{eqnarray}
\psi_{nm}({\boldsymbol{\xi}},\tau) & = & H_n\left(\frac{x}{2^{1/2}}\right) H_m\left(\frac{y}{2^{1/2}}\right) e^{\frac{\xi^2}{4}} \\\nonumber
& & \cdot e^{-i(n+m+1)\tau}
\end{eqnarray}
where $H_n(g)$ are the Hermite polynomials.

Since the initial superposition has a finite temperature, the initial mode probabilities are given by the linear, $\mu = 0$, Bose-Einstein distribution namely
\begin{equation}\label{equ:mb}
|a_{nm}|^2 = \frac{1}{e^{(n+m+1)/\beta}-1}
\end{equation}
where the scaled temperature $\beta = k_BT/\hbar\omega_t$.
The noninteracting thermal state has no long range order, this incoherence enables the mode phases of the initial state, $\alpha_{nm}$, to be described by a random variable on the interval $(0,2\pi)$.
The numerical simulation must then be run several times using different initial states to ensure that the kinetics are independent of the initial phases.

The SHO modes in the initial superposition are thermally distributed, however because of the mode cut-off numbers, $O,P$, the initial mean field is a truncation of a thermally-excited state and therefore, not in equilibrium.
The resulting truncation error, however, is negligible as the choice of initial parameters, $\beta = 20$ and $O,P = 20$, ensures that the initial distribution is sufficiently accurate, whilst still being computationally manageable.
Furthermore the nonlinear counterparts of the unpopulated SHO states, $\psi_{nm}$ $n,m \geq 20$, should not be unfavourably populated as all symmetry combinations initially exist.
Therefore the thermalisation process will couple all the states populating them irrespectively of the initial mode amplitude.

\begin{figure}
\begin{center}
\includegraphics[width=8cm,height=8cm]{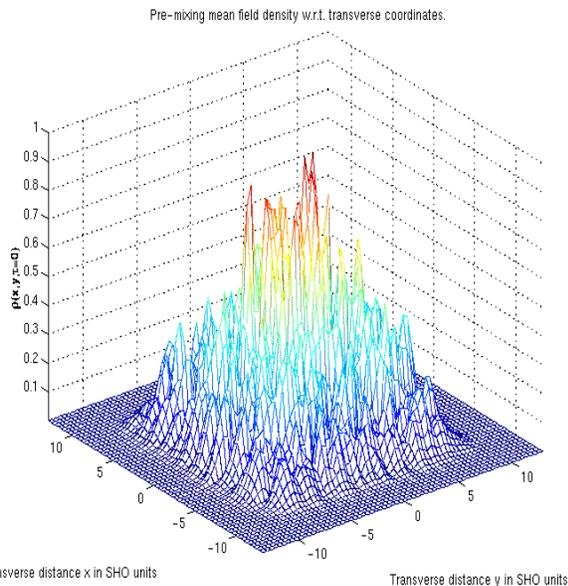}
\caption{Surface plot of the mean field density of a typical initial state, $\Psi_{\mathrm{I}}$, with respect to the scaled spatial coordinates $({\boldsymbol{\xi}},0)$ where $O<P + 20$, $\beta = 20$.}
\label{fig:one}
\end{center}
\end{figure}

The initial state, $\Psi_{\mathrm{I}}({\boldsymbol{\xi}},\tau)$, is then propagated using the method described in section \ref{sec:back}, where the interaction is adiabatically increased to its final value over the first half of the mixing time and the mean field is allowed to equilibrate over the following half.
The increase in the nonlinearity from zero, where the initial condition is an exact solution to the GP equation, to its final value corresponds to filling the trap with like particles, increasing the number of atoms in the thermally-excited state.

To obtain a link between the number of atoms in the thermally-excited state and its zero temperature counterpart the final value of $\eta$ was chosen such that $N$ was equivalent to the number of atoms in a Thomas-Fermi state, $N_{TF}$, where
\begin{equation}
N_{TF} = \int|\psi_{TF}({\boldsymbol{\xi}},\tau)|^2 {\mathbf{d}}{\boldsymbol{\xi}} = \frac{\mu^2}{4a}\left(\frac{\hbar}{m\omega_t}\right)^{1/2}
\end{equation}
Choosing $\mu = 515$ results in a total number of atoms $N = 3.07\times 10^7$.
Combining this number of atoms with a trap ratio $\epsilon = 4\times 10^9$ ensures that the nonlinear factor, $\eta$, is large enough to result in a nonlinear mixing rate that allows full thermalisation of the Bose-condensed gas.
The large numbers involved simply reflect the slow pace of the nonlinear mixing in two dimensions due to the reduced mode density.

The numerical method results in the thermally-excited state, $\Psi_{\mathrm{T}}({\boldsymbol{\xi}},\tau_m)$, given by
\begin{equation}
\Psi_{\mathrm{T}}({\boldsymbol{\xi}},\tau_m) = \left(\prod_{j=Q}^0 e^{-i \delta \tau\hat{H}}\right)\Psi_{\mathrm{I}}({\boldsymbol{\xi}},0)
\end{equation}
where $Q$ is the number of computational steps in the time dimension, $\delta \tau$ is the scaled step size and the scaled mixing time $\tau_m = Q \delta \tau$.
The simulation was run with $\tau_m = 100$, where the nonlinearity is increased on the interval $\tau_m = [0,20]$ and then remains constant allowing the mean field to equilibrate.
The spatial grid has 256$\times$256 computational points with a scaled area of 50$\times$50 and  the number of computational steps $Q = 20000$.
This results in a calculation that requires some hours to run on a Sun Ultra II workstation.

\begin{figure}
\begin{center}
\includegraphics[width=8cm,height=8cm]{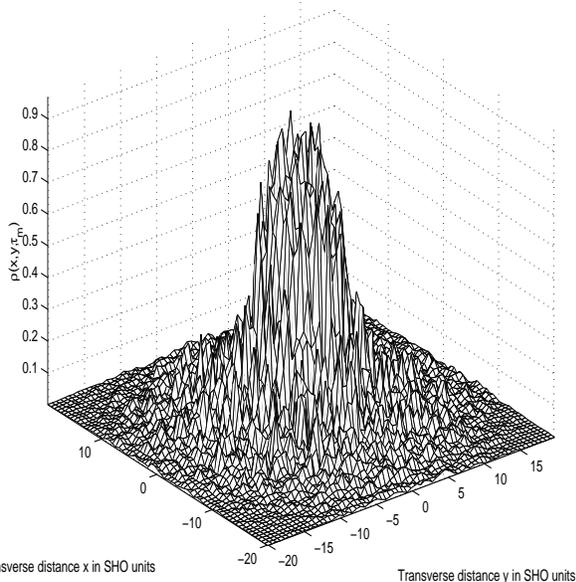}
\caption{Surface plot of the mean field density of a typical thermally-excited state, $\Psi_{\mathrm{T}}$, with respect to the scaled spatial coordinates $({\boldsymbol{\xi}},0)$ after a scaled mixing time $\tau_m = 80$.}
\label{fig:two}
\end{center}
\end{figure}

The results of the numerical simulation of the thermally-excited state generation are shown in 
Figs.\ \ref{fig:one} and\ \ref{fig:two} which show the density, $\rho({\boldsymbol{\xi}},\tau) = |\Psi({\boldsymbol{\xi}},\tau)|^2$, of the thermally-excited state before and after the nonlinear mixing process has occurred at $z=0$.
The figures show that the thermalisation leads to a macroscopically populated ground state, in the centre of the figure, surrounded by thermally-excited modes.
The geometry of the initial state, which could have been circular but was chosen to be square for didactic and diagnostic purposes, has been changed into a circular one reflecting the symmetry of the trapping potential.
This geometric relaxation of the mean field, into an equilibrium state, provides a visually convincing display of the nonlinear mixing.
The geometric relaxation also results in a lower temperature, a consequence of the two body scattering process acting to give the mean field the most probable distribution resulting in a maximal entropy state.

\begin{figure}
\begin{center}
\label{fig:three}
\includegraphics[width=8cm,height=8cm]{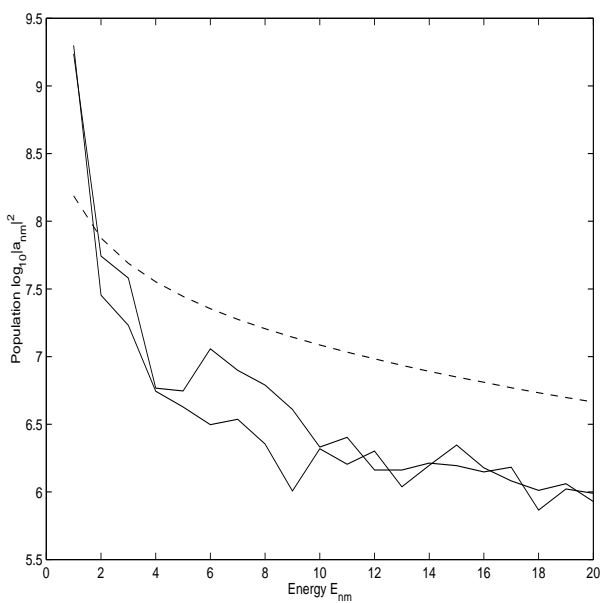}
\caption{Mode populations, $|a_{nm}|^2$, of the initial (dashed) and self-consistent thermally-excited states (solid) for two initial choices of the phase. It is plotted with respect to the scaled linear mode energy, $n+m+1$, after a scaled mixing time $\tau_m = 80$.}
\end{center}
\end{figure}

Further evidence of the thermalisation is shown in Fig.\ \ref{fig:three}, which shows the populations of the normal modes for the initial linear and final nonlinear states for two initial choices of the phase, calculated using the method of adiabatic following.
The population distributions demonstrate that the nonlinear mixing is sufficient to redistribute the energy of the system into an interacting Bose-Einstein distribution, in the equipartition sense.
This redistribution was found to occur if the initial population distribution was Boltzmann or Bose-Einstein, although the initial choice did effect the temperature of the resulting thermally-excited state.

Figs.\ \ref{fig:one}-\ref{fig:three} are typical of the results obtained, where the qualitative features of the final density, $\rho({\boldsymbol{\zeta}},\tau_m)$, and mode populations, $|a_{nm}(\tau_m)|$ described above are independent of the choice of initial phases.
This is shown in Fig.\ \ref{fig:three} and in subsequent simulations, even if the phase of the initial modes are chosen such that the initial state is coherent.

%%
%%
%% SECTION 4
\section{Evaporative Cooling}
\label{sec:evap}

With the ability to create thermally-excited states, a model of the evaporative cooling process can now be investigated.
The process of evaporative cooling involves the selective removal of the most energetic atoms in the thermally-excited state followed by a period of rethermalisation resulting in a net cooling effect.
In current experiments the cooling is carried out by inducing a radio frequency transition to an untrapped magnetic sublevel.
The transition frequency is a function of the magnetic field and this dependence results in the selective removal from the trap of atoms with a potential energy $E > \hbar\omega_{rf}|m_f|$ where $\omega_{rf}$ is the angular frequency of the transition inducing laser.
The frequency of the radiation is then gradually lowered bringing less energetic atoms into resonance with the untrapped state with the thermally-excited state rethermalising as the evaporative process continues.

This energy dependent removal of the atoms makes it possible to view the evaporative cooling either as an absorption of the edges of the condensate, or the lowering of the trap walls, allowing the escape of the more energetic atoms.
Of the two options, it is computationally simpler to introduce an absorbing boundary, the position of which is a function of time.
Such a model does not make any assumption or approximation other than representing the state by a classical field or that the cooling can be modelled by a spatially dependent removal of the field.

Inclusion of the absorption term, $A({\boldsymbol{\xi}},\tau)$, in the GP equation results in the time dependent nonlinear Schroedinger equation
\begin{eqnarray}\label{equ:hamc}
i\partial_t \Psi({\boldsymbol{\xi}},\tau) & = & \left(-\nabla^2 + \frac{\xi^2}{4} +\eta|\Psi({\boldsymbol{\xi}},\tau)|^2 \right)\Psi({\boldsymbol{\xi}},\tau) \\\nonumber
& & -i A({\boldsymbol{\xi}},\tau) \Psi({\boldsymbol{\xi}},\tau).
\end{eqnarray}

Since the absorbing boundary must have a smooth profile to preclude the introduction of computational errors, $A({\boldsymbol{\xi}},\tau)$ was modelled using a Super-Gaussian
\begin{equation}
\label{equ:abb}
A({\boldsymbol{\xi}},\tau) = A_0 \left(1 - e^{-(\xi/\xi_e(\tau))^{2m}}\right)
\end{equation}
Here $\xi_e(\tau)$ defines the position of the step function and $A_0$ determines the effectiveness of the cooling which is related to the magnitude of the field of the evaporative cooling laser.
The super-Gaussian defines the spatial region that undergoes cooling and ensures a smooth transition between the absorbing and non-absorbing regions.
The integer $m$ which determines the spatial width, ${\boldsymbol{\xi}}_s$, over which the function steps from $0$ to $A_0$ is set at $20$ throughout this paper, giving an estimated width of ${\boldsymbol{\xi}}_s = 0.2 \times {\boldsymbol{\xi_e}}(\tau)$.
This choice allows Eq.\ (\ref{equ:abb}) to rapidly step from $0$ to $A_0$ whilst still varying over a sufficiently large range to sample on a computational grid.

The evaporation length, ${\boldsymbol{\xi}}_e(\tau)$, implicitly determines the energy of the hotter atoms that are being removed from the thermally-excited state and as it is a function of time it also determines the dynamics of the cooling and ultimately the resultant ground state.

The cooling rate depends on the collision rate and therefore the density of atoms in the thermally-excited state and it is most effective, in terms of the resultant condensate population, if the absorbing boundary moves slowly removing only the minimum number of atoms.
To this end, two cooling strategies are compared, namely
\begin{mathletters}\label{equ:ablen}
\begin{eqnarray}\label{equ:ablena}
{\boldsymbol{\xi}}_e(\tau) & = & {\boldsymbol{\xi}}_i + ({\boldsymbol{\xi}}_f-{\boldsymbol{\xi}}_i)\tau/\tau_c \\ 
{\boldsymbol{\xi}}_e(\tau) & = & {\boldsymbol{\xi}}_ie^{(-\tau/\tau_s)} \label{equ:ablenb}
\end{eqnarray}\end{mathletters}
where ${\boldsymbol{\xi}}_i$ is the initial evaporation length;  ${\boldsymbol{\xi}}_f$ is the final evaporation length; $\tau_c$ is the time scale over which the cooling takes place and $\tau_s = -\tau_c/\ln\left(1-\xi_f/\xi_i\right)$.
The evaporation length has a simple linear variation in Eq.\ (\ref{equ:ablena}) and an exponential variation in Eq.\ (\ref{equ:ablenb}) although both have identical initial and final lengths.

Numerical simulations of the evaporative cooling process modelled by the absorbing boundary, with $\xi_e(\tau)$ defined by Eq.\ (\ref{equ:ablena}-\ref{equ:ablenb}), were carried out using the numerical method described in section \ref{sec:back}.
The cooled state, $\Psi_{\mathrm{C}}({\boldsymbol{\xi}},\tau_c)$, is given by 
\begin{equation}
\Psi_{\mathrm{C}}({\boldsymbol{\xi}},\tau_c) = \left(\prod_{j=Q}^0 e^{-i \delta \tau\hat{H}}\right)\Psi_{\mathrm{T}}({\boldsymbol{\xi}},0)
\end{equation}
where the cooling time, $\tau_c = Q\delta \tau$.
The number of computational steps and transverse grid size is the same as in section \ref{sec:exs} with the cooling time, $\tau_c = 80$, which results in a computation taking the same real time to complete as before.
The initial condition used is a thermally-excited state, $\Psi_{\mathrm{T}}({\boldsymbol{\xi}},0)$, created in the same manner as section \ref{sec:exs} with the numerical constants remaining unchanged.

Striking differences in the results of the evaporative cooling occur depending on the choice of $\xi_i$, two specific cases are considered below.
The results also show, however, that there is little variation between the exponential and linear dependence of the evaporation length, Eqs.\ (\ref{equ:ablen}).
Although the linear dependence, Eq.\ (\ref{equ:ablena}), results in a slightly larger ground state population and a slightly warmer condensate than the exponential dependence.
Subsequent figures presented in this paper relate to this linear case.

\subsection{Cooling: $\xi_i \sim$ size of the thermally-excited state.}\label{sec:cool1}

In this case the evaporation length varies monotonically from $\xi_e = 17$ to $\xi_e(\tau = 60) = 8$ during the cooling cycle followed by 20 scaled time units of equilibration resulting in $\tau_c = 80$.
The change is slow enough to ensure that only the most energetic atoms are removed from the thermally-excited state.
The resulting rethermalisation process redistributes the population of the normal modes of the G-P equation, reducing the temperature.
Typically the cooling cycle starts with a density, $\rho({\boldsymbol{\zeta}})$, as shown in Fig.\ \ref{fig:three} in the previous section and results in the density plot shown in Fig.\ \ref{fig:four}.
The evidence that the thermally-excited state has cooled is three fold:

\begin{figure}
\begin{center}
\includegraphics[width=8cm,height=8cm]{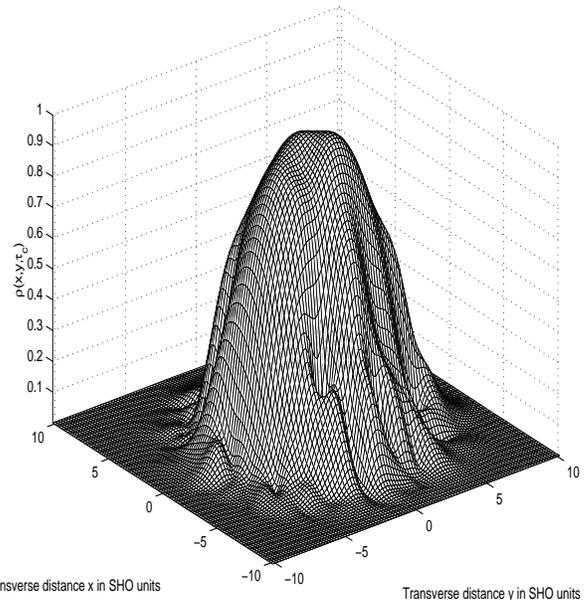}
\caption{Surface plot of the mean field density of an evaporatively cooled state, $\Psi_{\mathrm{C}}$, with respect to the scaled spatial coordinates $({\boldsymbol{\xi}},0)$ after a scaled cooling time $\tau_c = 60$ where $\xi_i >$ size of the thermally-excited state.}
\label{fig:four}
\end{center}
\end{figure}

\paragraph{} The density of the condensate is now much smoother and thus the velocity, ${\boldsymbol{v}}({\boldsymbol{\zeta}}) = \frac{-i\hbar}{2m\rho({\boldsymbol{\zeta}})}(\Psi^*{\boldsymbol{\nabla}}\Psi - \Psi{\boldsymbol{\nabla}}\Psi^*)$, of the initially thermally-excited state has been reduced.
This reduction reflects the fact that the thermally-excited state now has a lower temperature. Also the shape of the resulting condensate density has a profile with a greater similarity to that of the ground state of the GP equation.

\begin{figure}
\begin{center}
\includegraphics[width=8cm,height=8cm]{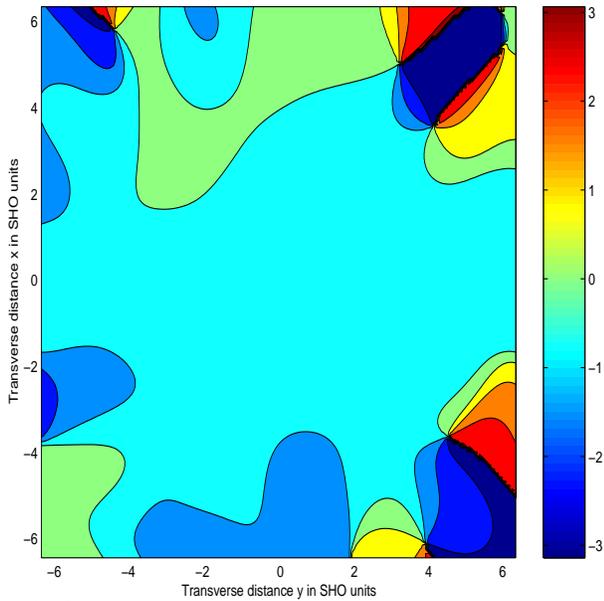}
\caption{Contour plot of the phase of the mean field of an evaporatively cooled state, $\Psi_{\mathrm{C}}$, with respect to the scaled spatial coordinates ${\boldsymbol{\xi}}$ after a scaled cooling time $\tau_c = 60$.}
\label{fig:five}
\end{center}
\end{figure}

\begin{figure}
\begin{center}
\includegraphics[width=8cm,height=8cm]{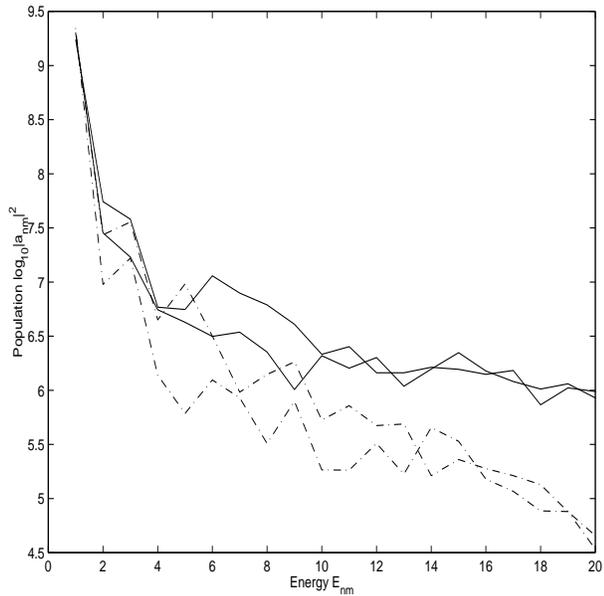}
\caption{Mode populations, $|a_{nm}|^2$, of the thermally-excited (solid) and cooled (dash-dot) states for two initial choices of the phase. It is plotted with respect to the scaled linear mode energy, $n+m+1$, after a scaled cooling time $\tau_c = 60$.}
\label{fig:six}
\end{center}
\end{figure}

\paragraph{} The phase of the thermally-excited state, Fig.\ \ref{fig:five}, is now much smoother and has large coherent areas. This arises because the thermally-excited state now predominantly consists of a ground state with only a small number of modestly populated excited states producing the phase fluctuations.

\paragraph{} Fig.\ \ref{fig:six} shows the populations of the normal modes of the thermally-excited state before and after cooling for two initial choices of the phase, calculated using the method of adiabatic following.
This figure clearly demonstrates that the thermally-excited state has been cooled.
The figure also shows that the atoms from the more energetic states have not just been removed, but that the population has been a transfered to the ground state as a result of the rethermalisation process.

\subsection{Cooling: $\xi_i <$ size of the thermally-excited state.}\label{sec:cool2}

\begin{figure}
\begin{center}
\includegraphics[width=8cm,height=8cm]{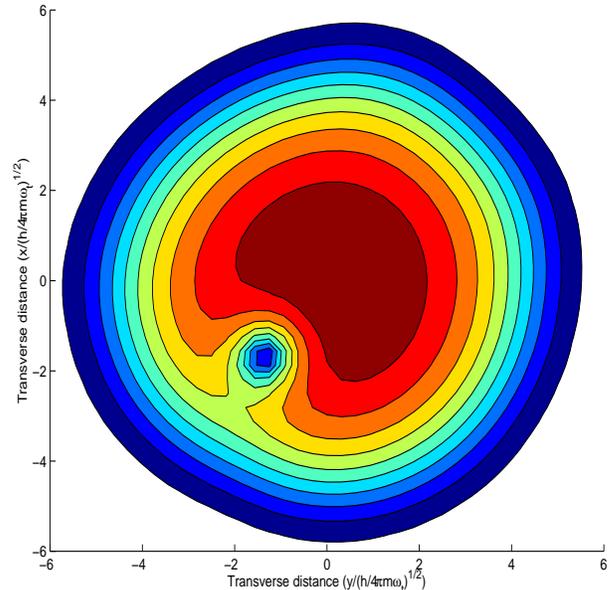}
\caption{A contour plot of the mean field density of an evaporatively cooled state, $\Psi_{\mathrm{C}}$, with respect to the scaled spatial coordinates $({\boldsymbol{\xi}},0)$ after a scaled cooling time $\tau_c = 60$ where $\xi_i <$ size of the thermally-excited state. The trapped vortex state is on the lower left-hand edge of the condensate.}
\label{fig:seven}
\end{center}
\end{figure}

In this case the evaporation length varies monotonically from $\xi_e = 12$ to $\xi_e(\tau = 60) = 7$ during the cooling cycle, giving an initially larger spatial cut into the thermally-excited state. This is then followed by 20 scaled time units of equilibration resulting in $\tau_c = 80$.
The initially larger spatial cut results in the removal of some atoms which may have a significant amount of kinetic energy.
Cooling in this fashion typically results in a mean field density, shown in Fig.\ \ref{fig:seven}, which includes a trapped vortex state.

Initially the thermally-excited state has no angular momentum, so when viewed in an angular momentum representation the vortex states exist in an equal number of right and left handed pairs.
Therefore, when the cooling is carried out as described in section \ref{sec:cool1} the vortex pairs annihilate one another exactly.
If, however, $\xi_i$ is less than the size of the thermally-excited state, the initially deeper spatial cut into the thermally-excited state removes a vortex during the initial stages of the cooling cycle resulting in an unbalanced vortex population.
Consequently, as the thermally-excited state is cooled, at least one vortex will remain trapped with the ground state, resulting in a state with a non zero angular momentum.

The resulting trapped vortex state is robust and remains with the condensate until the spatial confinement, due to the evaporative cooling, causes the state to be energetically unfavourable at which point it disappears.

This method of vortex trapping may be directly applicable to two-dimensional experimental situations if the evaporation surface in the experiment can be setup so that it removes atoms with a proportion of kinetic energy in the initial stages of cooling.

%%
%%
%% SECTION 6
\section{Conclusion}
\label{sec:conc}

We have shown in this paper that using the GP equation to time evolve an initial state is a numerically efficient method for the creation of self consistent thermally-excited states. 
Results presented demonstrate that the GP equation acts to thermalise an initially non-equilibrium state through two body scattering processes described by the nonlinearity within it.
The resulting nonlinear mixing generates a self-consistent thermally-excited state with an interacting Bose-Einstein population distribution, in the equipartition sense.
Where the ensemble average of the GP equation describes the finite temperature kinetics and dynamics of a boson gas if and only if the quasiparticle populations $|a_{nm}|^2 \gg 1$.

A model of the evaporative cooling process without the assumption of sufficient ergodicity or approximations relating to the strength of the interactions between the thermal and condensate parts of the thermally-excited state was presented.
This model uses an extension of the GP equation based on the removal of the most energetic atoms in the spatial distribution.
Results from a two-dimensional numerical simulation show that the evaporative cooling model produces a visibly cooler mean field density and a population transfer to the ground state, which gives rise to a coherent condensate.
These results demonstrate that the model accurately reflects the evaporative cooling process and consequent rethermalisation of the thermally-excited state.
Furthermore other results presented show that under a different cooling regime vortex states can be trapped along with the condensate and it is noted that it may be possible to achieve this using current experimental configurations.

The possibility of creating self-consistent thermally-excited states and the numerical simulation of these states in two dimensions allows the study of many new and interesting phenomenon such as vortex and other quasiparticle interactions, coherence properties of thermally-excited states and more accurate models of Bose-condensation experiments.

\section*{Acknowledgements}

The authors acknowledge the support of the United Kingdom Engineering and Physical Sciences Research Council.

%%
%%
%% REFERENCES

\end{multicols}

%%
%%
%% FIGURES

\end{document}